\documentclass[10pt,letterpaper]{article}
\usepackage{opex3}
\usepackage{amsmath}
\usepackage{cite}

\begin{document}

\title{Experimental generation of multi-photon Fock states}

\author{Merlin Cooper,$^{*}$ Laura J. Wright, Christoph S\"{o}ller, and Brian J. Smith}

\address{Clarendon Laboratory, University of Oxford, Parks Road, Oxford OX1 3PU, UK}

\email{$^{*}$m.cooper1@physics.ox.ac.uk} 



\begin{abstract}
We experimentally demonstrate the generation of multi-photon Fock states with up to three photons in well-defined spatial-temporal modes synchronized with a classical clock. The states are characterized using quantum optical homodyne tomography to ensure mode selectivity. The three-photon Fock states are probabilistically generated by pulsed spontaneous parametric down conversion at a rate of one per second, enabling complete characterization in 12 hours. 
\end{abstract}



\section{Introduction}
Individual photons are the fundamental quantum excitation of the electromagnetic field. Fock, or photon-number states, which are eigenstates of the number operator, are the generalization of single-photon states to higher excitation numbers. The ability to create Fock states of a well-defined single electromagnetic field mode with a prescribed excitation number is not only of foundational interest, but also has significant practical impact in quantum technology applications where Fock states are a key resource in the generation of single and multimode quantum states \cite{Holland:93, Banaszek:97, Pan:00, Kim:02, Asboth:05, Bencheikh:07, Bimbard:10, Bartley:12}. Many of these quantum applications require single-mode Fock-state sources that can be interfered in an optical network with coherent states and squeezed states to create highly entangled quantum states \cite{Loock:07}. A key requirement of these sources is the well-defined mode structure of the emitted state to enable high-visibility interference between different sources as well as auxiliary states.

In the last decade we have witnessed the first demonstration of optical homodyne tomography of a conditionally-prepared single-photon state \cite{Lvovsky:01,Zavatta:04,Huisman:09} and more recently a two-photon state \cite{Ourjoumtsev:06,Zavatta:08}. Furthermore, coherent superpositions of $|0\rangle$ and $|1\rangle$ photons \cite{Resch:02,Lvovsky:02}, and $|0\rangle$, $|1\rangle$ and $|2\rangle$ photon states \cite{Bimbard:10} have been prepared in the laboratory and characterized by means of optical homodyne tomography. More recent proposals \cite{Lance:06} and experimental demonstrations \cite{Ourjoumtsev:07} of Schr\"{o}dinger cat generation using $n$-photon Fock states and continuous-variable post-selection exemplify the importance of reaching into higher-order Fock layers for quantum state engineering. 

Here we demonstrate a major step in this direction by experimentally creating one-, two-, and three-photon states in well-defined ultrashort pulsed modes based upon heralded spontaneous parametric down conversion \cite{Mosley:08}. The source is well-matched to an auxiliary coherent state that can be used to generate entanglement at a beam splitter \cite{Kim:02, Asboth:05} and subsequently used to produce non-classical quantum states using conditional detection \cite{Bimbard:10, Bartley:12}. The well-timed nature of the source enables multiple sources to be concatenated with high-visibility interference. Furthermore, the source does not require narrowband spectral filtering of the down-converted photons \cite{Mosley:08}, thus enabling unprecedented brightness. This represents a significant milestone in quantum state engineering and detection, both in terms of the brightness of quantum states investigated and also the techniques used to characterize them.

\section{Experimental setup}
\begin{figure}
\centerline{\includegraphics[width=13cm]{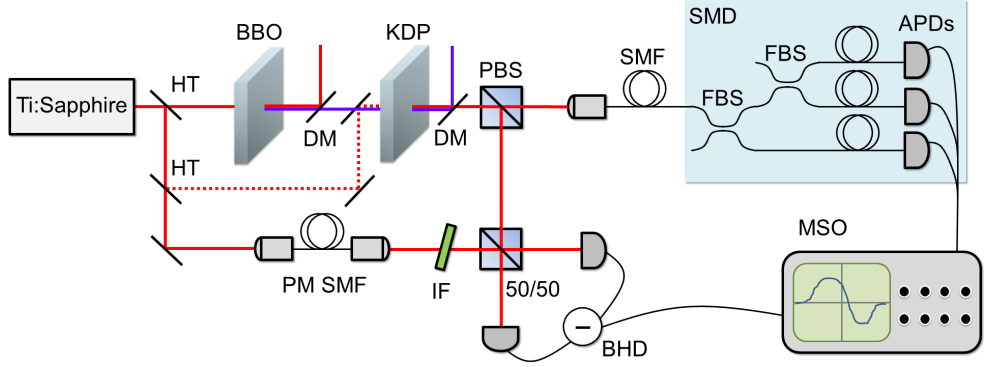}}
\caption{Schematic of the experimental setup for heralded Fock state tomography. \textbf{BBO}: beta barium borate crystal, \textbf{KDP}: potassium dihydrogen phosphate crystal, \textbf{PBS}: polarizing beam splitter, \textbf{HT}: highly-transmitting beam splitter, \textbf{DM}: dichroic mirror, \textbf{SMD}: spatially multiplexed detector, \textbf{FBS}: fiber beam splitter, \textbf{MSO}: mixed-signal oscilloscope, \textbf{BHD}: balanced homodyne detector, \textbf{SMF}: single-mode fiber, \textbf{PM-SMF}: polarization-maintaining single-mode fiber, \textbf{IF}: interference filter, \textbf{50/50}: 50/50 beam splitter, \textbf{APD}: avalanche photodiode.}
\label{fig:schematic}
\end{figure}

To create Fock states in a well-defined mode of the electromagnetic field, we generate a two-mode squeezed-vacuum (TMSV) state by the process of Type-II spontaneous parametric down conversion (SPDC) in a bulk potassium dihydrogen phosphate (KDP) crystal \cite{Mosley:08}. The SPDC source is engineered to allow the direct generation of pure photonic Fock states, without the need for tight spectral filtering in the trigger mode \cite{Mosley:08}, ensuring only a two-mode squeezed state is produced. The field generated is approximately described by
\begin{equation}
|\psi\rangle\approx\sqrt{1-\lambda^2}\left(|0,0\rangle + \lambda |1,1\rangle + \lambda^2 |2,2\rangle + \lambda^3 |3,3\rangle + O(\lambda^4)\right), \label{eq:TMSV_approx}
\end{equation}
where $\lambda$ is the squeezing parameter for the particular optical modes collected, and $|n,m\rangle$ is a two-mode state with $n$ ($m$) photons in the signal (trigger) mode. The intrinsic photon-number correlations between the two modes, in our case distinct polarization modes, enable the preparation of Fock states in one mode, which we shall call the signal, upon projection onto photon number in the conjugate mode, which we shall call the trigger mode.

A schematic of the experimental setup is shown in Fig.~\ref{fig:schematic}. An 80~MHz train of 100~fs pulses at 830~nm central wavelength from a Ti:Sapphire oscillator (Spectra Physics Tsunami) is frequency doubled with a conversion efficiency of 30\% in a 700~$\mu$m long $\beta$-barium borate (BBO) crystal resulting in 85~fs pulses at 415~nm. The residual fundamental is filtered out using dichroic mirrors and Schott glass filters, resulting in a maximum second-harmonic power of 800~mW. The second-harmonic beam is subsequently spatially filtered with a pinhole and focused into an 8 mm thick KDP crystal cut for Type-II collinear degenerate down conversion. A small fraction of the original 830~nm oscillator output is split off to serve as the local oscillator (LO) for homodyne detection of the heralded Fock states. We select a single spatial mode of the LO using a short (15~cm) polarization-maintaining single-mode fiber (PM-SMF), and shape the spectral mode by tilting an interference filter (IF) (Semrock LL01-830-12.5). The fiber length is chosen to be as short as possible so as not to significantly chirp the 3~nm LO pulse. The down-converted signal and trigger fields are orthogonally polarized and can therefore be separated at a polarizing beam splitter (PBS). The broader bandwidth trigger mode is coupled into a single-mode fiber and detected with avalanche photodiodes (APDs), thus heralding Fock states in the signal mode, which have a bandwidth of approximately 3~nm (duration on the order of 300~fs). Note, although we select a single spatial mode in the trigger arm using a single-mode fiber, it is not necessary to spectrally filter the mode since the SPDC source produces a spectrally uncorrelated two-mode state \cite{Mosley:08}.

One-, two-, and three-photon states are heralded in the signal mode by detecting exactly one, two, and three photons respectively in the corresponding trigger mode. Such photon number projections require a photon number resolving detector, i.e. a detector described by a positive-operator value measure (POVM) which is the set of projectors onto photon number states $\{ \hat{\Pi}_n = |n\rangle \langle n| \}$. Since such a detector is not readily available we instead use a pseudo-number-resolving detector by employing a spatially-multiplexed detector (SMD) comprising an array of three APDs (Perkin-Elmer SPCM-AQ4C) and two 50:50 fiber beam splitters (FBS). An incoming wave packet in the trigger mode is probabilistically split between the three APDs. The signal mode is then prepared in a one-, two-, or three-photon Fock state conditioned on obtaining exactly one, two, or three clicks respectively from the SMD. 

To characterize the quantum state of the signal mode generated for a given number of clicks in the trigger SMD, we employ a balanced homodyne detector. Using a high-bandwidth time-domain balanced homodyne detector (BHD) the electric field quadratures of the heralded Fock states are measured. Since the Fock states are not produced on demand, the BHD must be capable of distinguishing the laser pulses containing the heralded state from surrounding pulses which contain thermal light. This requires that the detector have an electronic bandwidth sufficiently greater than the pulse repetition rate of the source of 80~MHz. Additionally, the detector should exhibit low electronic-noise and excellent stability, in order to adequately distinguish the quantum noise of the state from other technical sources of noise. Our custom-designed BHD exhibits these necessary specifications and is described in detail elsewhere \cite{Cooper:12}.

\section{Fock-state tomography}
Characterization of the heralded Fock states using balanced homodyne detection has several advantages over other techniques such as direct photon counting using time-multiplexed detectors (TMDs) \cite{Achilles:03} or superconducting transition-edge sensors (TES) \cite{Hadfield:09}. Homodyne detectors can exhibit an efficiency in excess of 90\% whilst maintaining an electronic bandwidth sufficient for operating an experiment at 80~MHz or above, in contrast to TESs or TMDs which are limited to a range of kHz to a few-MHz repetition rates respectively. Operating at a high repetition rate is desirable since it allows for sufficient data collection of the rare probabilistic events in a shorter time, thus avoiding long-term experimental drifts. Not only does the use of a BHD allow higher efficiency and repetition rate compared with many photon counting detectors, but also more importantly ensures that only a single spatial-temporal mode, defined by the strong local oscillator, is examined.

Ideally, a photon source should produce Fock states in a well-defined, pure spatial-temporal mode, which is capable of exhibiting good interference with other sources in the same mode. Optical homodyne detection addresses a well-defined single-mode of the electromagnetic-field through the interference of an intense local oscillator (LO) with the unknown state in the matched signal mode. The interference visibility between signal and LO thus serves as an effective measure of the overlap between the optical modes, and results in an effective efficiency, $\eta_{\text{mm}}=V^2$, where $V$ is the classical interference visibility \cite{Aichele:02}.

For the non-classical Fock states under investigation in this work, there is no direct way to measure $V$. In order to achieve the initial overlap between LO and signal we are able to seed the down-converter with a pick-off from the original 830~nm oscillator beam, shown as the dashed red line in Fig.~(\ref{fig:schematic}). When this seed beam is properly overlapped with the blue pump inside the KDP, difference frequency generation (DFG) results allowing access to a bright alignment beam in a mode approximately the same as the conditionally prepared Fock states \cite{Aichele:02}. However, the DFG field can never be completely identical to the SPDC due to effects such as gain induced diffraction \cite{Kim:94} and the inability of the coherent field to exhibit perfect interference with a signal in a mixed quantum state. Therefore, the fine alignment between LO and signal is carried out directly with the heralded states, with the seed beam blocked. This is achieved by continuously acquiring pulses containing the heralded single-photon state from the homodyne detector and calculating the pulse amplitude variance compared to vacuum. The maximum overlap is achieved when the triggered pulse variance compared to vacuum is largest. 

Spatial overlap between the signal and LO is optimized and measured by coupling both modes into a single-mode fiber (SMF). The heralding efficiency into this fiber is defined as
\begin{equation}
\eta_{\text{he}}=\frac{{R_{\text{C}}}}{\eta_{\text{apd}} R_{\text{trigger}}},
\end{equation}
where $\eta_{\text{apd}}$ is the APD efficiency, $R_{\text{C}}$ is the raw coincidence rate and $R_{\text{trigger}}$ is the raw singles rate in the trigger mode. By ensuring near-perfect coupling of the LO into this overlap fiber, the measured heralding efficiency then serves as a direct measure of the degree of spatial overlap between the LO and signal-mode photon. We commonly achieve a heralding efficiency in excess of $\eta_{\text{he}}=0.65$ after correcting for $\eta_{\text{apd}}=0.45$.

\begin{figure}
\centerline{\includegraphics[width=13cm]{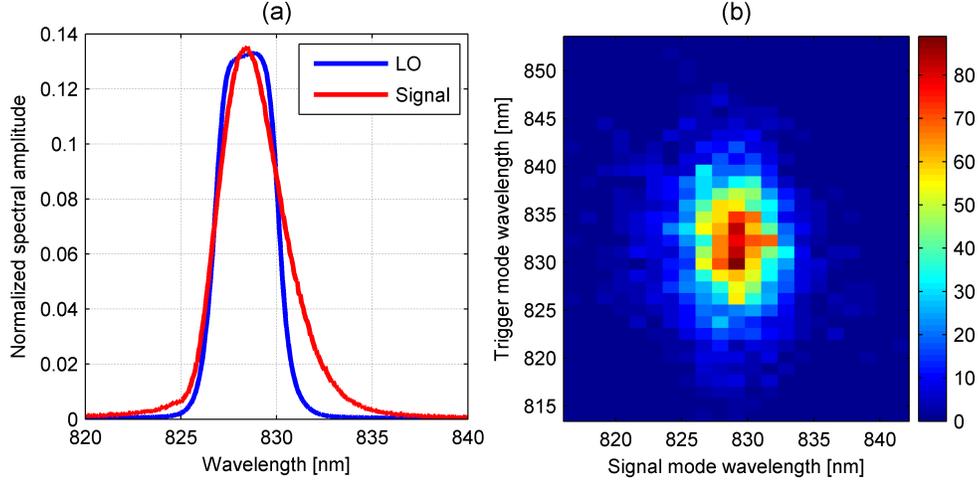}}
\caption{(a) LO (blue) and signal (red) spectral amplitudes, the calculated spectral overlap is 0.97. (b) Joint spectral intensity distribution for the collected signal and trigger modes from KDP crystal. The colour scale indicates the number of recorded coincidence counts for each spectral bin in the 2D scan. The calculated spectral purity of the state is 0.95.}
\label{fig:spectral_overlap}
\end{figure}

The spectral amplitude overlap is determined by measurement of the individual spectra of the LO and signal modes using an Andor Shamrock 303 spectrograph together with an Andor iDus 401 series CCD camera. The LO and signal spectra are shown in Fig.~\ref{fig:spectral_overlap}(a). The two spectra exhibit an overlap of 0.97, assuming a flat spectral phase. 

A joint spectral intensity (JSI) measurement of the trigger and signal modes collected from the KDP crystal was performed in order to estimate the spectral purity of the heralded state. To perform this measurement, single-mode fibers collecting the two modes of the down-converter were sent to two separate spectrographs and the outputs were collected by scanning multi-mode fibers \cite{Kim:05}. This enabled a spectrally-resolved coincidence measurement of the two modes, i.e. a JSI measurement, as shown in Fig.~\ref{fig:spectral_overlap}(b). The number of Schmidt modes present in the collected two-mode state, assuming a flat phase distribution, then allows determination of the state purity \cite{Mosley:08}. We obtain an estimate of $P=0.95$ for the state purity. This confirms the SPDC source is generating a spectrally uncorrelated state in the spatial modes under investigation, and thus we do not require any spectral filtering in the trigger mode.
\begin{figure}
\centerline{\includegraphics[width=13cm]{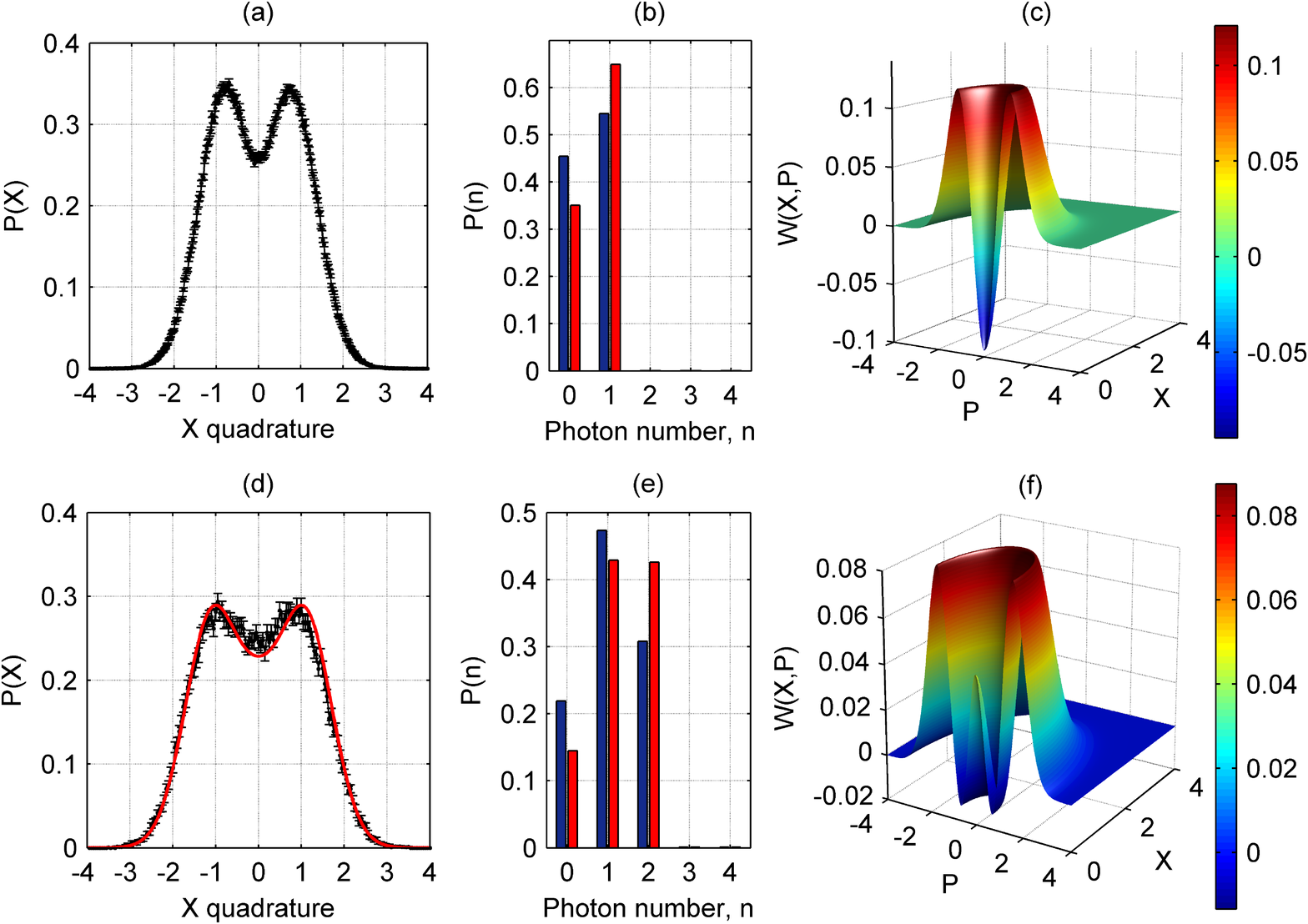}}
\caption{Marginal distributions $P(X)$ of the heralded (a) single- and (d) two-photon states, showing clear non-Gaussian profiles. Photon number statistics $P(n)$: raw (blue) and corrected for the detector efficiency $\eta_{\text{bhd}}$ (red) for (b) single- and (e) two-photon states. Wigner functions $W(X,P)$ of the reconstructed (c) single- and (f) two-photon states, both corrected for the detector efficiency $\eta_{\text{bhd}}$.}
\label{fig:single_photon}
\end{figure}

To optimize the overlap of the signal and LO we first performed quantum state tomography (QST) of a heralded single photon. The homodyne detector output was recorded using a Tektronix MSO5104 oscilloscope triggered by a single APD in the trigger mode of the down converter. Production rates for the single-photon state in excess of $180,000~\text{s}^{-1}$ were routinely observed, thus allowing real-time computer-controlled optimization of parameters such as the temporal overlap of the signal and LO. In general the reconstructed state, neglecting higher-order terms, will be an admixture of vacuum and single-photon components
\begin{equation}
\hat{\rho}_{\text{meas}}=(1-\eta)|0\rangle\langle 0 |+\eta |1\rangle\langle 1 |,
\end{equation}
where $\eta$ is the overall homodyne efficiency for the measurement. This comprises four main contributions: 1) the detector efficiency $\eta_{\text{bhd}}$, 2) the mode-matching between LO and the conditionally-prepared single photon $\eta_{\text{mm}}$, 3) the non-unit purity of the heralded state $\eta_{\text{p}}=\sqrt{P}$ \cite{Grosshans:01}, and 4) detector dark counts leading to false-positives, $\eta_{\text{dc}}$.  $\eta_{\text{bhd}}$ was determined to be 0.85 by performing tomography of well-calibrated coherent state probes. $\eta_{\text{mm}}$ is estimated to be 0.66 based on the measured heralding efficiency and spectral overlap between LO and the heralded state and the JSI measurement gives an estimate of $\eta_{\text{p}}=\sqrt{0.95}=0.97$. Dark counts are insignificant since they represent a background of only $300~\text{s}^{-1}$ on a trigger rate of $180,000~\text{s}^{-1}$. This gives an overall estimate of $\eta_{\text{est}}=\eta_{\text{bhd}}\eta_{\text{mm}}\eta_{\text{p}}\eta_{\text{dc}}=0.54$.

The raw homodyne data were inverted using a maximum likelihood algorithm \cite{Lvovsky:04} implemented in MATLAB to obtain the density matrix of the detected state. The raw reconstructed state is consistent with a pure single-photon being subject to an efficiency of $\eta=0.545$. This value is clearly in excellent agreement with the estimated value of $\eta_{\text{est}}=0.54$. The marginal distribution $P(X)$, photon number statistics $P(n)$ and Wigner function $W(X,P)$ of the detected state, conditioned on the one-click event are shown in Fig.~\ref{fig:single_photon}(a-c). The single-photon Wigner function, which is corrected for the detector efficiency $\eta_{\text{bhd}}=0.85$, is strongly negative at the origin of phase space, $W(0,0)=-0.095$.

To herald two- and three-photon Fock states, the trigger mode of the down-converter is coupled to a multi-channel spatially-multiplexed detector, in place of the single APD used to herald the single-photon state. The oscilloscope trigger logic is set to acquire homodyne data only when two or three of the APDs in the SMD click simultaneously. For each trigger event we contemporaneously acquire pulses containing only vacuum in order to calibrate the detector. We observe production rates of approximately 200~s$^{-1}$ ($1~\text{s}^{-1}$) for the two- (three-) photon state, thus allowing acquisition of a sufficient data in about 5 minutes (12 hours). When performing tomography of the three-photon state we utilized the maximum available 415~nm pump power of 800~mW, whereas only a fraction was used for producing the one- and two-photon states in order to minimize higher-order terms. The squeezing parameter $\lambda$ in Eq.~(\ref{eq:TMSV_approx}) is estimated to be 0.071 during the one- and two-photon state experiments and 0.087 for the three-photon experiment, when utilizing the full pump power. 

For the heralded two-photon state we acquired 60,000 quadrature data samples in 5 minutes, while for the heralded three-photon state 40,000 quadrature samples were acquired over a period of 12 hours. It is interesting to note that this is roughly the same time taken to acquire data for the original single-photon homodyne tomography experiment, one decade previous to this work \cite{Lvovsky:01}. This demonstrates the improvement in both photon sources and detector technology, giving rise to brighter Fock-state sources and high-bandwidth homodyne detection. This avoids the need to use amplified laser systems and pulse-pickers, thereby significantly reducing experimental complexity. The measured marginal distribution for the two-photon state and corresponding reconstructed photon number statistics and Wigner function are shown in Fig.~\ref{fig:single_photon}(d-f). The measured marginal distribution $P(X)$ is consistent with the predicted distribution, shown as the red curve in Fig.~\ref{fig:single_photon}(d), which we determine by calculating the marginal distribution of a pure two-photon state subject to preparation with overall efficiency $\eta=0.545$. Furthermore, the two-photon state Wigner function exhibits negativity after correction for the detector efficiency $\eta_{\text{bhd}}=0.85$.

\begin{figure}
\centerline{\includegraphics[width=13cm]{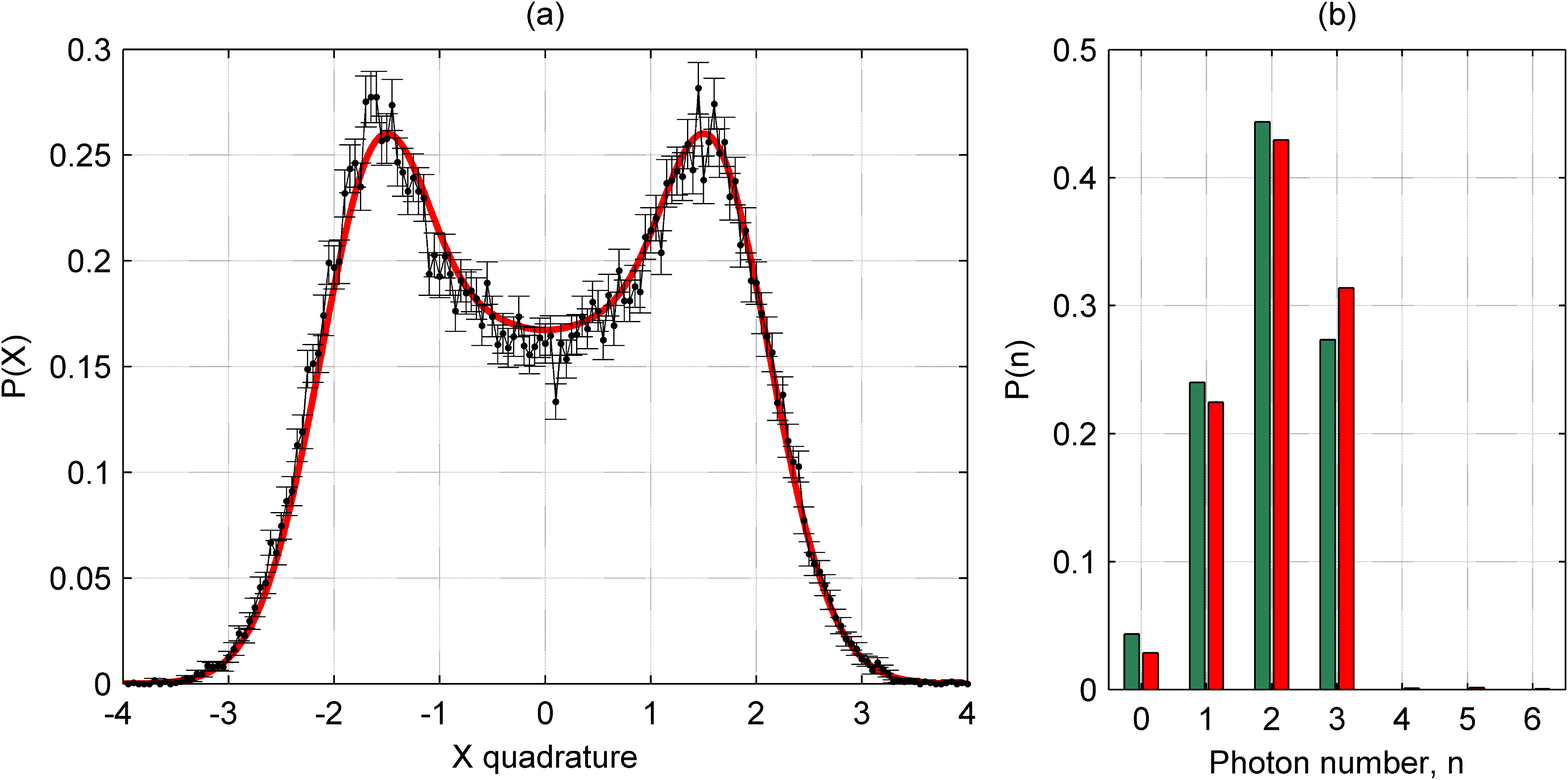}}
\caption{(a) Predicted (red line) and measured (black dots) marginal distribution $P(X)$ with error bars for the three-photon state. (b) Photon number statistics $P(n)$ for predicted (green) and reconstructed (red) state, both corrected for detector efficiency $\eta_{\text{bhd}}$.}
\label{fig:three_photon_marginal}
\end{figure}

Figure~\ref{fig:three_photon_marginal}(a-b) shows the measured marginal distribution and photon number statistics of the heralded three-photon state. The predicted raw marginal distribution is calculated by subjecting an ideal three-photon state to the overall efficiency, $\eta=0.545$, which was determined from the single-photon reconstruction. The assumption made is that the overall detection efficiency due to mode-matching and the detector efficiency is the same for the single-photon and three-photon state. It is evident that the measured quadrature statistics of the pulse containing the heralded three-photon state correspond well with the predicted marginal, with small errors introduced most likely due to the limited digitizer resolution of 8-bits. The error bars are calculated based on the expected uncertainty, $\sqrt{N}$, for each quadrature bin containing $N$ events. Thus the bin size is chosen so as to show enough detail in the marginal distribution whilst ensuring statistical errors are not overly significant.

\begin{figure}
\centerline{\includegraphics[width=13cm]{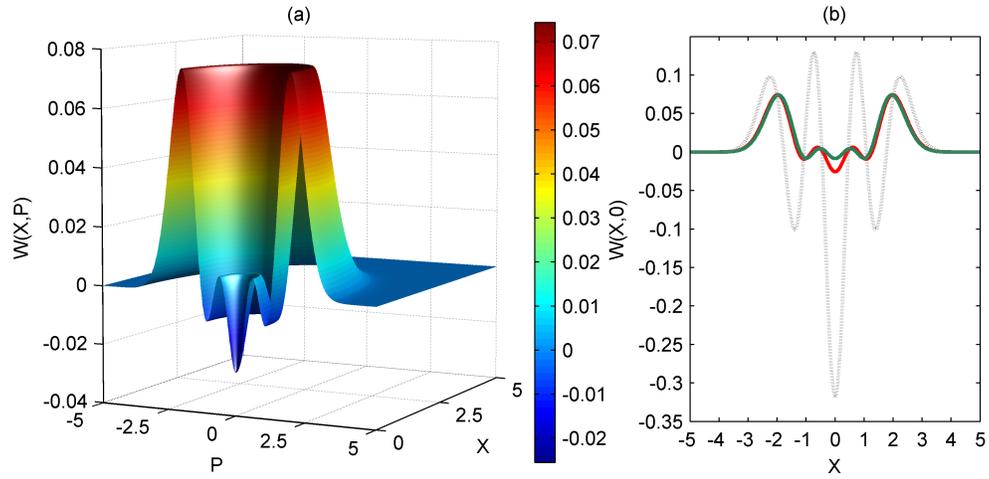}}
\caption{(a) Wigner function $W(X,P)$ of the reconstructed three-photon state corrected for detector efficiency $\eta_{\text{bhd}}$, showing negativity around the origin of phase space. (b) Cross-section of Wigner function in $P=0$ plane: reconstructed state (solid red line), predicted state (solid green line) and perfect three-photon state (dashed gray line).}
\label{fig:three_photon}
\end{figure}

As with the one- and two-photon states, the measured quadrature statistics are inverted using a maximum likelihood algorithm, correcting for the detector efficiency of $\eta_{\text{bhd}}=0.85$. Fig.~\ref{fig:three_photon_marginal}(b) shows a plot of the reconstructed photon number statistics $P(n)$ for both the predicted state (green) and the reconstructed state (red). We find an excellent correspondence between the predicted and measured states and suspect the discrepancies are partly due to higher order terms in the measured state. The Wigner function of the state reconstructed from the homodyne data, after correcting for the detector efficiency, is presented in Fig.~\ref{fig:three_photon}(a) and exhibits negativity near the origin of phase space, as seen in the cross section, Fig.~\ref{fig:three_photon}(b). We calculated the fidelity between the predicted and reconstructed three-photon state defined as
\begin{equation}
F=\text{Tr}\left[\left(\sqrt{\hat{\rho}_m}{\hat{\rho}}_p\sqrt{\hat{\rho}}_m\right)^{1/2}\right],
\end{equation}
where $\hat{\rho}_m$ and $\hat{\rho}_p$ are the density matrices of the measured and predicted states respectively. The fidelity is found to be 99.7\% which represents a remarkable correspondence between the experimentally reconstructed thee-photon state and the predicted state, thus giving further credence to characterization of such states by homodyne tomography.

\section{Conclusion}
In conclusion, we have shown the first experimental demonstration of broadband heralded multi-photon Fock states occupying well-timed single-mode wave packets. Characterization by means of optical homodyne tomography ensures that a single mode is examined. This work will serve as a building block for experimental investigations of quantum coherence across multiple modes and higher-order Fock layers, with great promise in quantum information applications. Future work to improve the spatial mode overlap of the heralded photons and increase the brightness of the two-mode squeezed state should enable higher preparation efficiency and photon-number states to be realized.

\section*{Acknowledgments}
We are grateful for helpful discussions with M. Barbieri and M. Karpi\'{n}ski. This work was supported by the University of Oxford John Fell Fund and EPSRC grant No. EP/E036066/1.


\end{document}